%% file: sample-manuscript.tex
\documentclass[manuscript,screen]{acmart}
\usepackage{colortbl}
\usepackage{listings}

\AtBeginDocument{%
  \providecommand\BibTeX{{%
    \normalfont B\kern-0.5em{\scshape i\kern-0.25em b}\kern-0.8em\TeX}}}

\acmConference[CDLT Technical Report 2023-01]{Centre for DLT Technical Reports}{January 10, 2023}{Malta}
\acmBooktitle{Centre for DLT Technical Report 2023-01,
  January 10, 2023, Malta}

\input{solidity.tex}

\begin{document}

\title{Active External Calls for Blockchain and Distributed Ledger Technologies}
\subtitle{Debunking cited inability of Blockchain and DLT to make external calls}


\author{Joshua Ellul}
\authornote{Both authors contributed equally to this research.}
\orcid{0000-0002-4796-5665}
\author{Gordon J. Pace}
\authornotemark[1]
\orcid{0000-0003-0743-6272}
\affiliation{%
  \institution{Centre for DLT, University of Malta}
  \country{Malta}
}
\email{joshua.ellul@um.edu.mt}

%\author{Blind Author}
%\orcid{0000-0000-0000-0000}
%\author{Blind Author}
%\orcid{0000-0000-0000-0000}
%\affiliation{
  %\institution{Blind Institution}
  %\country{Blind Country}
%}
%\email{blind@blind.blind}

\renewcommand{\shortauthors}{Ellul and Pace}

\begin{abstract}
Blockchain and other distributed ledger technologies have enabled peer-to-peer networks to maintain ledgers with an immutable history and guaranteed computation, all carried out without the need of trusted parties. In practice, few applications of blockchain are closed i.e. do not interact with the world outside the blockchain, and various techniques have been proposed and used to handle such interaction. One problem is that it is widely accepted that, due to the decentralised nature of blockchain networks and constraints to ensure trust and determinism, such communication can only flow into the blockchain, and that blockchain systems cannot initiate and execute calls to external systems or services. In this paper we show that this misconception is preconceived by building on our previously presented solution to demonstrate that such calls can be directly initiated from the blockchain itself in a feasible and efficient manner. 
% \gp{Maybe add something on how we do this}
%\todo{It is widely accepted that blockchain and Distributed Ledger Technology (DLT) systems cannot execute calls to external systems or services due to each node having to reach a deterministic state. However, in this paper we show that this belief is preconceived by demonstrating a method that enables blockchain and distributed ledger technologies to perform calls to external systems initiated from the blockchain/DLT itself.}
\end{abstract}



\maketitle

\section{Introduction}
\label{sec:introduction}

%It is probably the case, that Bitcoin \cite{nakamoto2008peer}, a `peer-to-peer electronic cash system', hit the headlines because of the manner in which it enabled peer-to-peer payments in a decentralised and trustless manner. 
% JE -- initially I don't think this made it popular, its first claim to fame was probably drugs, so probably better that we do not claim specifically why

%Over the past decade, Bitcoin \cite{nakamoto2008peer}, a `peer-to-peer electronic cash system', hit the headlines many times for different reasons from Forbes's `crypto currency' article\footnote{\url{https://www.forbes.com/forbes/2011/0509/technology-psilocybin-bitcoins-gavin-andresen-crypto-currency.html}}, to it being used to buy pizza\footnote{\url{https://bitcointalk.org/index.php?topic=137.0}}, and various periods of hype surrounding related cryptocurrency activities \cite{ellul2021blockchaindead}. However, the science behind the scenes which enabled such a system is, interesting on its own right. A few years after Bitcoin's proposal and launch, Ethereum~\cite{wood2014ethereum} showed that the approach can be extended further to enable smart contracts --- user built applications that execute in a decentralised yet trustless manner. Guarantees are provided seamlessly so long as blockchain transactions and smart contracts are executed within the boundaries of the blockchain. Things start becoming more complex when interaction outside the blockchain is required. 

Bitcoin~\cite{nakamoto2008peer}, a `peer-to-peer electronic cash system', triggered interest in the design and use of trustless transaction systems. In particular, a few years after Bitcoin's proposal and launch, Ethereum~\cite{wood2014ethereum} showed that the approach can be extended further to enable smart contracts --- user built applications that execute in a decentralised yet trustless manner. Guarantees are provided seamlessly so long as blockchain transactions and smart contracts are executed within the boundaries of the blockchain. Things, however, start becoming more complex when these processes require interaction with entities outside the blockchain. 

With public permissionless blockchains, incoming interaction (such as a user invoking certain smart contract functionality, or an oracle notifying a smart contract of the temperature being read by a sensor), is standard practice and supported without any issues.\footnote{Here we are limiting ourselves to how such communication is achieved. Needless to say, there is the issue of trust in the oracle providing the information, which is beyond the scope of this paper.} On the other hand, in many cases, it would be desirable to have smart contracts actively initiate external calls\footnote{The term \emph{external call} in the context of smart contracts has been used for both calls from one smart contract to another (on the same blockchain) and to calls from a smart contract to functionality outside the blockchain. In this paper we use the term exclusively for the latter.} to fetch data as part of their computation. For instance, a smart contract computing the cost for a pay-per-use of a rented space, may need to query an external trusted system to fetch the number of units of electricity consumed over the time period in question\footnote{One could assume that the electricity provider may provide an API to retrieve such data.}. If one were to na\"ively allow such outgoing queries to external systems, two problems arise. Firstly, miners attempting to execute the transaction to add it to a new block, or verifying a new block proposed by another miner, would all send the request to the external party\footnote{As was proposed in Hyperledger composer: \url{https://hyperledger.github.io/composer/v0.19/integrating/call-out}} --- potentially flooding it with requests. If these external requests have an effect on the real-world, this may be even more worrying or render the approach unsuitable --- think of an external call to \emph{`raise the thermostat temperature by 1 degree' being performed by multiple miners.}. Secondly, certain queries may be non-deterministic, resulting in the external service sending different responses to the different miners validating a block, thus making it impossible for them to validate the response reported by the miner who constructed the block.

% \je{I feel we are making too strong a connection to the physical world (IoT) which may throw off some reviewers, it may make more sense to add change an example to a more typical use case of blockchain --- I'll work on this.}

Both problems arise due to such external calls possibly having real-world side-effects. For these reasons, it has long been claimed and accepted that supporting external calls from smart contracts on a permissionless public blockchain is impossible to achieve.

The solution typically adopted in order to address this limitation is that of using oracles --- external third parties making offchain data available onchain, whether through the oracle regularly pushing data onto the blockchain (e.g. sending the rate of exchange at regular intervals), or by having the oracle monitor the blockchain state and react whenever a request for external information is written onchain, upon which event the oracle will provide the response (e.g. a user requests the value of a particular sensor in a particular format). 

In practice, the former solution is not always possible, since fulfilling the request for data may require dynamic input. Even if not, however, the solution comes at a substantial expense, since a continuous stream of external data is to be written to the blockchain whether or not some smart contract will use it. 
%Another shortcoming of this solution is that if the external call is intended, not to get input from the real world, but to cause a real-world effect e.g. trigger an actuator to open a door, this solution does not work. 
%JE: this is calling it an external call, but its not an external call, its input from the external world.
The second solution requires the request to be written on the blockchain before a response is provided, thus spreading computation over multiple blocks which may not be practical (particularly if a computation consists of a chain of external inputs with each one depending on the results of previous ones) and does not allow for programming inline use of the external data. 

Furthermore, we note that both solutions may require another point-of-trust --- the oracle --- over and above the service from which the input is required. Indeed, techniques to determine the veracity of whether the data originates from a particular trusted party are often used (mostly using signed results) but the dependence on this relaying third party still remains.

We have in a short paper \cite{ellul2022verifiable} (and pre-print \cite{ellul2021towards}) presented the initial approach demonstrating that active external calls can be supported by blockchains (including public ones) whilst avoiding the problems mentioned above. The proposed solution allows for miners (or validators) to actively make external calls by requiring services to respond with signed input which will be written together with the associated transaction. In this paper, we further evaluate the approach through demonstration of a typical use-case to assess overheads when compared with a standard oracle approach, and demonstrate that the approach is not only feasible from a gas consumption perspective (as previously demonstrated \cite{ellul2022verifiable}), but also with respect to overall blockchain transaction throughput (for use-cases requiring external inputs). We also, through the use-case, discuss potential lower of smart contract code complexity when using the proposed approach.
% \gp{Should we expand this paragraph?}

%and then having them write the result together with the evidence of it veracity. We have evaluated this approach on a use case to assess the overheads when compared to a standard approach, and we have showed that there are substantial gains. \gp{Should we expand this paragraph?}

The paper is organised as follows. In Section~\ref{sec:motivation} we motivate the problem in detail, and present existing state-of-the-art. We then present our solution in Section~\ref{sec:solution} and evaluate it with respect to the prevalent existing approach in Section~\ref{sec:eval}. We finally conclude in Section~\ref{sec:conc}.

% \todo{
% The paper is organised as follows. In Section~\ref{sec:motivation} we motivate further the need for direct calls emanating from public blockchains and present the technical challenges which have stood in the way of building solutions to this problem till now. We then explain the solution we are proposing in the remaining sections. Section~\ref{sec:ver} explain the encoding of verifiable external calls, Section~\ref{sec:tx} presents the extended structure of transactions to handle them and its impact on blocks in Section~\ref{sec:blk}. Finally, Section~\ref{sec:net} brings these together at the level of the network.
% }

\section{Motivation and Challenges}
\label{sec:motivation}

% Blockchain and DLT-based systems require that the decentralised logic encoded within them reaches a deterministic state. It is said that every node must execute the exact same logic in order to achieve consensus.

With any digital technology which goes beyond data and information processing, communication with the real world in order to read and update the state of real-world objects is a crucial element. With the key benefit of blockchain-based systems being the decentralisation of trust, this is particularly salient. As long as all the logic and information lies within the blockchain-platform, such as pure cryptocurrency transfers, no centralised trust issues arise, but the moment we need data from the outside world e.g. requiring external exchange price-pairs, or we need to query or trigger devices in the real world e.g. unlock a door, we introduce new points of centralised trust: the exchange, the device unlocking the door, and potentially an in-between relayer interfacing between the blockchain platform and the external service or device. More notably, recent trends in the blockchain space, such as DeFi (Decentralised Finance) depend heavily on such external trusted party data \cite{caldarelli2021blockchain}.

\emph{Oracles}~\cite{DBLP:journals/access/BreikiRSS20} are typically used to bridge between on-chain computation and the off-chain world. Although the term is normally used to refer to intermediaries allowing smart contracts to access data from the outside world, it has also been used for intermediaries enabling actuation within the off-chain world ~\cite{DBLP:conf/bpm/MuhlbergerBFCWW20}. Whether the interaction is incoming or outgoing, and whether it is in the form of a push (initiator triggers) or a pull (receiver fetches), oracles are an additional layer of trust. In addition, to make the role of such oracles sustainable, they have to be adequately incentivised, raising further issues of trust.

External data fed into a blockchain by oracles who actively initiate transactions requires the oracles to pay for providing information. Many oracles that could provide useful information may not be incentivised to do so --- why should an external party provide information at a cost if there is nothing in it for them? Solutions to this problem have been proposed which require external systems and networks to feed such oracle input into smart contracts \cite{al2020trustworthy} --- again, however, requiring that such costs are borne, typically by the end-users or dApp (decentralised application) operator. Also, feeding in oracle input imposes delays, some which may be due to the oracles' own delays in submitting required transactions, and others due to the fact that different steps within an interaction will take place in different blocks.

It would be ideal if oracles can be directly accessed by blockchains and that any relayers in-between can be bypassed, allowing blockchain and DLT systems to actively make calls to external systems and, if need be, incentivise the carrying out of such calls within the internal economics of the blockchain. However, it is the general consensus in the community that such direct external calls are not possible~\cite{greenspan2016many,weber2016,ellis2017chainlink,xu2017,rimba2017,lu2017adaptable,marchesi2018agile,adler2018astraea,xu2018,shae2018,worley2018,van2018publicly,gatteschi2018,carminati2018,victor2018,grootemanproviding19,molinabenefits19,ma2019reliable,kamiya2019shintaku,guarnizo2019pdfs,da2019trustable,dinh2019blueprint,guarnizo2019pdfs,shi2019blockchain,ibba2019agile,cappiello2019,van2019,liu2019,daniel2019,caitruth20,muhlberger2020foundational,Lafourcade20,lo2020,Beniiche20,mammadzada2020,woo2020,caldarelli2020,ladleif2020,woo2020distributed,wang2020,nelaturu2020,zhao2020,lv2021,milani2021,chen2021,marchesi2021,basile2021,zhao2022towards,lin2022survey,mekouar2022survey}.

\subsection{The Challenges}
\label{sec:technicalchallenges}
For small private DLT systems, a solution to initiate externals calls to trusted parties had been proposed\footnote{https://hyperledger.github.io/composer/v0.19/integrating/call-out} which requires that each node makes the same call to the external system and receives the exact same response. While works, albeit in an inefficient manner, for small networks, the solution does not scale up with network size. Moreover, this solution limits the types of calls that can be made to services that always will provide the same response irrespective of (i) when the call is made; and (ii) how many times it is made: they have to be deterministic and referentially transparent. The reason is that, for such a mechanism to work, all nodes querying the external system must receive the same response, including at a later point in time when a block is being verified. Using the solution on an external service lacking this property would limit the extent to which the chain could be verified if an external service provider is either no longer available or returns different responses to the original response received. More so, it would be impossible in future to determine whether the participating verifying nodes actually recorded the correct response, or whether they had all agreed on an incorrect response. Whilst approaches for non-interactive deterministic communication have been proposed\footnote{https://ethresear.ch/t/on-chain-non-interactive-data-availability-proofs/5715} \cite{adler2019building}, to the best of our knowledge the work presented herein is the first proposal for active external calls which still allows for network-wide consensus regarding computed state whilst also not requiring all nodes to execute the same external call.

So far, dApps require that parties which interact with the blockchain network must initiate the communication themselves (or make use of proxy services to do it on their behalf e.g. see ~\cite{zhang2016town}). This is not ideal as it limits parties: (i) to those that are willing to interact and integrate with the specific smart contract/network and bear any costs required to initiate transactions; or (ii) by requiring dApps to make use of external services which become trusted parties themselves. Ways of mitigating trust issues with such parties have been proposed. For instance, Chainlink~\cite{ellis2017chainlink} avoids having trust concentrated in a single node oracle by using decentralised nodes, and provides a bridge between various data sources and blockchain systems --- yet it requires fees to be paid for providing the service\footnote{\url{https://medium.com/@chainlinkgod/scaling-chainlink-in-2020-371ce24b4f31}}.

\begin{figure}
\centering
\includegraphics[scale=0.85]{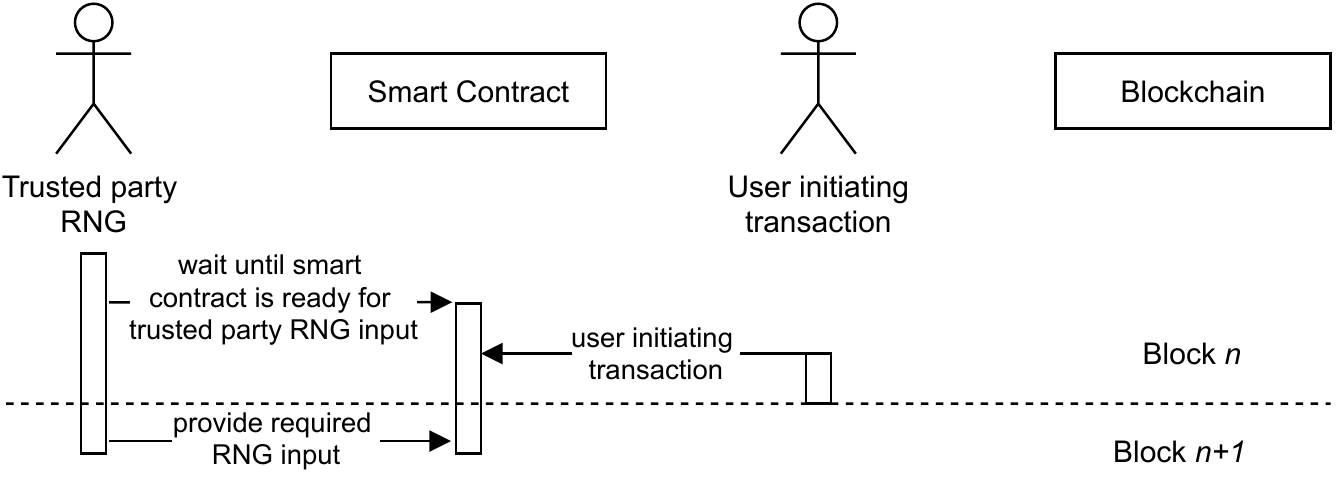}
\def\svgwidth{\columnwidth}
\caption{Trusted parties may often require to wait for an action in a first block before it can transmit its input into a subsequent block.}
\label{fig:rng}
\end{figure}

Furthermore, input coming from trusted parties would result in separate transactions in different blocks, which could result in potentially large delays for processes to carry out computation making use of such oracles. Each dependency on such external parties would have to be split into three phases: the request for the data coming from the external party is recorded, the external party records its input, and finally the computation proceeds using that data. Delays can be large when trusted parties are actively monitoring the ledger, and can be even larger if the trusted external parties are not listening for such changes in real-time. 

Consider random number generation (similar to that provided by Provable,\footnote{\url{https://provable.xyz}} previously called Oraclize) in a decentralised betting dApp. One could go about different ways to implement this, such as by using a commit and reveal pattern~\cite{wohrer18} or using an external oracle to provide the random number. In a process similar to that described by~\citet{oraclizerng}, the trusted random number input should only be revealed after the smart contract process reaches a certain stage --- for instance, if the bet is to guess the random number, this should not be revealed before the user places the bet. If a trusted-third party oracle is used, the transaction initiating  the user's bet would take place in block $n$, with the input from the random number generator oracle taking place in block $n+1$ or later. Figure~\ref{fig:rng} depicts a sequence diagram depicting the sequence of events. After that, the rest of the logic to resolve the bet can proceed. This could be implemented as a continuation of the transaction providing the random number, or triggered in a later block i.e. in block $n+2$ or later. As this example shows, there is a need for the request and response to span multiple blocks. It is worth noting that this bypasses the issue of determinism and referentially transparent computation since the information is initiated from outside the blockchain.

\subsection{Contribution Overview}
If smart contracts could directly and actively call oracles to acquire key input to a dApp, many of the issues discussed could be circumvented. However, the general consensus is that it is impossible or infeasible to allow blockchain systems to make such active external calls to external services due to: (i) the requirement for computation to reach a deterministic state~\cite{sankar2017survey}; and (ii) potential overloading of points-of-trust with requests. Also, whenever a node requires to verify some computation which involves a point-of-trust, it would have to send another request, possibly getting a different result than what the original caller received, and furthermore inundating the centralised node with more requests (potentially resulting in a denial-of-service attack). 

\begin{figure}
\centering
\includegraphics[scale=0.90]{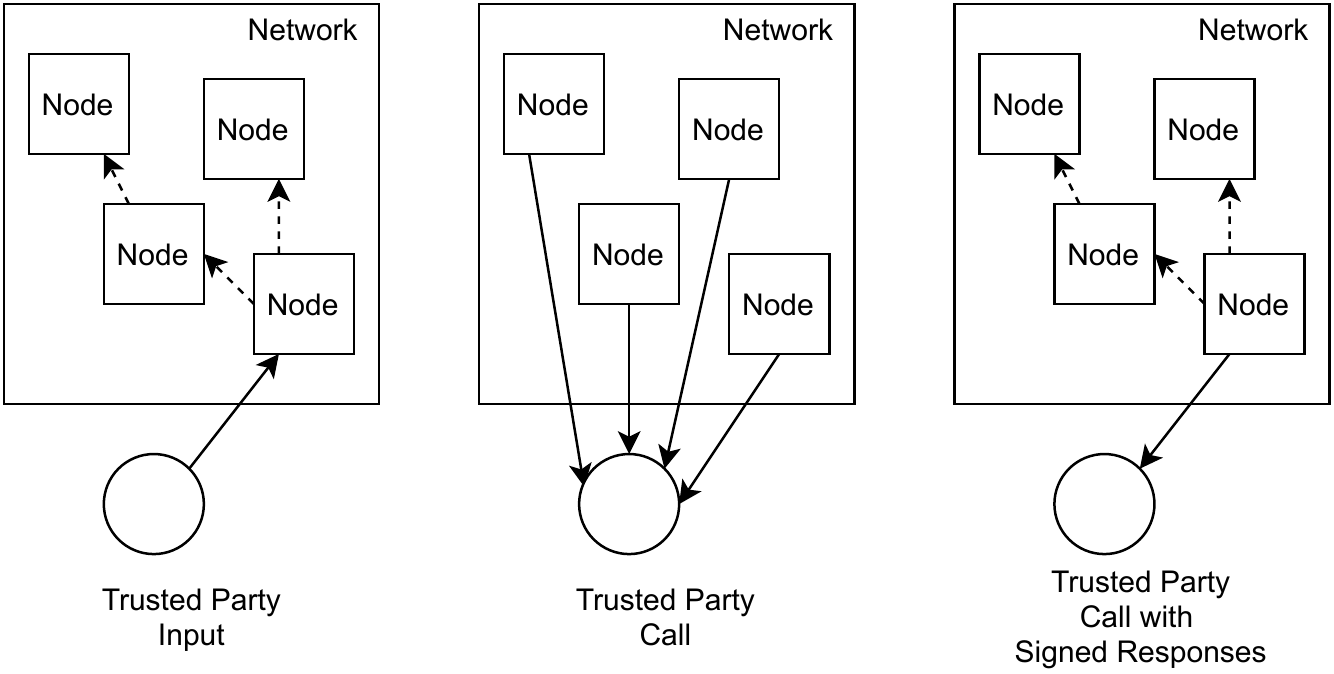}
\caption{Left: traditional trusted party input; middle: active calls requiring each node to undertake the external call that must return the same input; right: external calls enabled with verifiable signed responses. Reproduced from \cite{ellul2022verifiable}}
\label{fig:signedresp}
\end{figure}

In this paper we present a solution circumventing these issues enabling the implementation of a blockchain platform supporting such external calls using
%Firstly we note that Perhaps based upon the often cited deterministic nature of computation that is required --- yet whilst this statement is true, it is important to highlight that it is the state which computation reaches that must be deterministic, and the computation performed can reach such a deterministic state in different ways. In this paper 
a mechanism that allows for such systems to interact with external parties directly in a feasible manner. In Figure~\ref{fig:signedresp}, an overview of the oracle input transaction/call flow is provided, with distinctions between three distinct modalities: (i) the conventional oracle input, depicted on the left; (ii) external calls that are inefficient due to the requirement for deterministic responses to be made and received by all nodes, displayed in the center; and (iii) our proposed methodology of utilising verifiable external calls, shown on the right.

\section{Design and Implementation}
\label{sec:solution}
We will now specify the concrete requirements of the problem at hand, and present our solution, explaining how the requirements are addressed through our algorithm.

% \todo{We will now provide design bla bla bla}

\subsection{Verifiable External Calls}
\label{sec:ver}
The solution we present enables the use of verifiable external calls i.e. requests (calls) made within a smart contract to an external system that returns back a signed response, and which: (i) can be verified to truly be a response from the external party in question without having to repeat the queries; (ii) which does not require any further communication (with the external party or any other party). Such assurances can be provided in a similar manner to that done in a traditional approach, by verifying whether the response was digitally signed by the expected party. It is essential that knowledge of the trusted party's public key is known, whether through prior knowledge (as is often the case with oracle input, where knowledge of the trusted party's address is necessary) or through some other trusted entity or registry. 

To allow for blockchain code (and smart contracts) to make external calls directly in a manner that is feasible and efficient, we propose verifiable external calls which provide guarantees with respect to the veracity of the origin of the response both at the time of processing and at any point in future. 

We define a verifiable external call as a tuple including: a request, a public key and a signed response structured as follows: $\langle request, \;public\_key,\; signed\_response \rangle$.

The \emph{request} would typically be represented as a Uniform Resource Identifier (URI) and should reference the external system to be called (which may also comprise of other input data). The \emph{public\_key} may be hard-coded into the application logic or retrieved by a trusted certificate registry. Whether the \emph{public\_key} had been hard-coded or retrieved prior to the call being made, the \emph{public\_key} requires to be recorded along with the response data, since other nodes will need to verify that the response originated from the respective external party. The \emph{signed\_response} is the response that has been signed using the private key associated with the aforementioned \emph{public\_key}.

Indeed, this proposed approach requires that the trusted data sources that will be queried provide an end-point that will respond back with a digitally signed response. However, recent proposals and implementations indicate that such infrastructure may eventually be adopted as a standard\footnote{\url{https://wicg.github.io/webpackage/draft-yasskin-http-origin-signed-responses.html}\\ \url{https://developers.google.com/web/updates/2018/11/signed-exchanges}}. If adopted,  this approach would allow for the blockchain to call any external service (that is up-to-date with such standards) requiring no further work to facilitate integration.

Furthermore, to avoid play back of old responses, a nonce or some other form of challenge-response could be made use. Whilst, challenge data sent to the external party will be part of the \emph{request}, the verifiable external call's definition may be extended to include the challenge-response. For example, the request can be augmented by a request number or a fresh nonce $\nu$, which is expected to be included unchanged in the response:
%\small\[ \langle request \oplus \{\mathit{request\_nonce}\mapsto\nu\}, \;public\_key,\; signed\_response\oplus \{\mathit{response\_nonce}\mapsto\nu\} \rangle \]\normalsize

\centerline{$\langle request \oplus \{\mathit{request\_nonce}\mapsto\nu\}, \;public\_key,\; signed\_response\oplus \{\mathit{response\_nonce}\mapsto\nu\} \rangle$}

\subsection{Transactions}
\label{sec:tx}
When a transaction is initiated (be it by a user, another system, or the system itself if the underlying DLT allows it) and accepted for execution, the node which is processing the transaction will establish all external calls which need to be performed, execute them and record the responses received back from the external parties along with associated digital signatures. However, for this not to induce a denial-of-service attack on the service provider, the process has to be split into separate steps:

% Indeed, at this point, miners must ensure that the response is from the expected external party by verifying the response and signature against the trusted party's public key. Furthermore, if a unique number, timestamp, or challenge-response mechanism was used to ensure that old data is not repeated, then this would also be validated at this point. 

\begin{enumerate}
    \item A transaction containing an external call (be it due to a smart contract, a native transaction, or otherwise) goes to the miner's mempool --- we will call it an \emph{initiator transaction}.
    \item A miner intending to process such an initiator transaction will see the need for an external call. They \emph{do not} perform the external call, but simply include the intention to execute it as part of the block they are constructing (with any unique nonce to ensure results from the external source are not repeated).
    \item Once a miner solves a block, they perform the external call append it as an extension to block (not included in the hash calculation) before broadcasting it to the network.
    \item Verifiers of a block with such transactions must verify that the result (with with any accompanying unique nonce) is indeed signed by the external service.
\end{enumerate}

\begin{figure}
\centering
\includegraphics[scale=0.9]{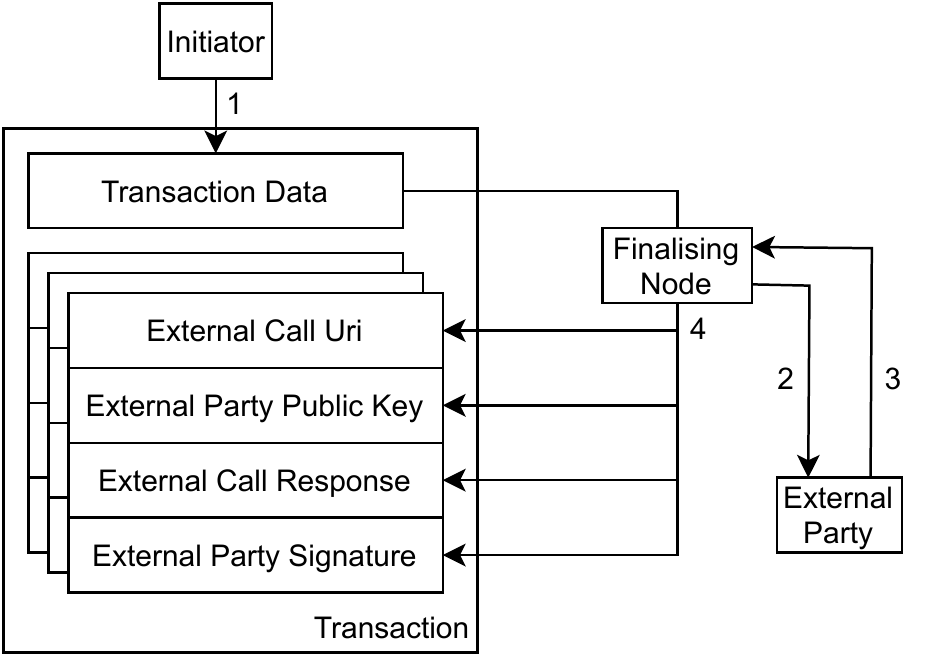}
\caption{Transaction finalisation process. Reproduced from \cite{ellul2022verifiable}.}
\label{fig:trans}
\end{figure}

Figure~\ref{fig:trans} demonstrates how a transaction is initiated and further associated with the external call response data.

If any responses are not verified, then the transaction may be deemed to have failed, or depending upon reparation or other logic the transaction may still be valid (and able to process the unverified response). This is a design decision that each platform would need to consider. The same goes for external calls for which no response is received. 

There are a number of issues to address to ensure that the proposed solution works as claimed.

\begin{itemize}
    \item Since adding external calls to a block will result in the miner delaying (albeit in a small manner) the broadcast of a discovered block, there is the question of why they would bother including such transactions in their blocks. Adequately large transaction fees for such transactions would entice miners to include them, and appropriate mechanisms to correlate the cost between normal transactions and ones containing external calls would have to be in place. This is not much different than the correlation between gas costs and transaction fees for native transfer transactions.

    \item The first issue naturally raises a second one. Why would a miner not perform the external transactions before discovering a block in order to be able to minimise the time between discovering a block and broadcasting it? If miners were to behave in this manner, there would be a denial-of-service attack on the service with all miners attempting the calls at once. In practice, given that external calls are substantially expensive to carry out, it may not make sense for a miner to perform them just in case they are required. However, one can piggy-back on the signature mechanism in order to address this issue --- by requiring the hash of the first part of the block to be sent to the service and be included in the signed response. This could act as a nonce, and would ensure that the calls can only be done \emph{after} the (first part of the) block is found. Obviously, this means that not all services can be used off-the-shelf for external calls, but this equally holds if a counter or timestamp is used to avoid replay attacks as was proposed earlier. 
    
    \item Finally, we need to ensure that participating nodes execute such external calls rather than simply record that such a call failed, requiring an incentive or assurance mechanism. Verifiable external calls could also be undertaken by the transaction initiator (i.e. the party submitting the transaction can provide the information from the external call as part of the transaction submission process, but this depends upon the architectural design of the blockchain, smart contract and wallet/dApp software submitting the transaction. By performing the verifiable external call at transaction submission time (on the initiator), the aforementioned problem pertaining to nodes potentially reporting back failed external calls would be eliminated.
\end{itemize}

Another modification to our approach to address these issues is to split the main block and external call results to be broadcast separately, thus separating the two phases of block definition. This would require an additional layer of complexity to handle situation where the second part of the a block is not received by a portion of the network miners. These issues require further investigation, but we will focus on investigating the proposed solution in the rest of the paper.

% To reiterate, to ensure that a miner does not repeat old responses from external parties, the response and signature could be accompanied with the date and time the response was generated and/or a unique response identifier associated with the response (and potentially request as well). 
\subsection{Implementation Details}
\label{sec:impl}
We reproduce the implementation details from 
\cite{ellul2022verifiable} here for ease of reference:
 \begin{quote}
   The Go Ethereum (geth) node implementation (version 1.16.5) was modified to include support for the verifiable external call mechanism described above. The following salient modifications were implemented.\\

   \textbf{EXCALL Transaction}
A new type of transaction, an EXCALL transaction (in \textbf{excall\_tx.go}), was added (on top of the existing Legacy and Access List transactions) to facilitate storing the additional data associated with external calls (described in Section~\ref{sec:ver}) in an EXCALL tuple --- containing the external call response, signature, and the external party's known public key.\\

\textbf{EXCALL instruction}
A new \texttt{EXCALL} virtual machine opcode which instructs the virtual machine to execute the external call was added. Rather than modify the whole programming tool-chain (including the Solidity programming language and Solidity compiler) to support the proposed \texttt{EXCALL} instruction, for the purpose of this prototype it was emulated by replacing \texttt{PUSH32} instructions (used for string assignments) whose associated data starts with ``http'' into \texttt{EXCALL} ones.\footnote{Indeed, this means that in the prototype it is not possible to make use of a \texttt{PUSH32} instruction for data that starts with the string ``http'', however this does not impact the prototype's purpose to evaluate the proposed technique.}\\

A miner executes the emulated \texttt{EXCALL} instruction only when finalising a block, and will undertake an external call to the URL specified as a parameter to the instruction. Upon receiving a response and a valid digital signature for the respective public key, the relevant data will be appended to an EXCALL transaction.\\

Following this, the transaction is stored in the block with the EXCALL transaction data filled in. This then allows for other nodes to verify the external call based upon the stored data (without having to initiate an external call itself).
 \end{quote}

\section{Evaluation}
\label{sec:eval}
To evaluate the approach we have analysed different blockchain networks to identify a common use-case requiring external oracle data and found that requiring random number input is a popular use of oracles. We have built a use-case for both (i) a standard oracle implementation; and (ii) the active external call approach being proposed in this paper. We evaluate the two in terms of: (i) aspects pertaining to the complexity of code required to implement the use-case; and (ii) experimental evaluation of achievable throughput for the different implementations.

\subsection{The use-case}
\label{sec:usecase}
A betting dApp similar to that discussed in Section~\ref{sec:technicalchallenges} is being used to  serve the purpose of a required use-case to evaluate performance of the proposed approach against a traditional approach. However, the evaluation discussed below would also apply to other smart contract use-cases that have similar protocol requirements where a party must first initiate a transaction to a smart contract prior to external oracle data being made available on the blockchain (which typically is due to not wanting to reveal that data prior to the initiating transaction).

The use-case, in a manner similar to that depicted in Figure~\ref{fig:rng}, requires that these steps are followed to complete a betting transaction:

\begin{enumerate}
    \item A user initiates interaction with the smart contract placing a bet, sending an amount of cryptocurrency to the smart contract.
    \item Data from the oracle is retrieved and fed into the smart contract to determine whether the user won.
\end{enumerate}

Note that we leave the smart contract game-agnostic, in that rather than having the external party provide a random number to decide the outcome of the bet, we simply rely on that party to decide whether the punter won the bet. For instance, in case of a fair coin toss game, the oracle would be expected to (fairly) decide on a 50:50 basis whether the punter won.\footnote{The user need not even choose heads or tails, since either way they have a 50\% chance of winning.} On the other hand, in case of a game to guess a number between 1 and 100, the oracle would (fairly) decide whether the player won on a 1:99 basis.

The use-case has been implemented in Solidity for: (i) a standard Ethereum network; and (ii) a modified Ethereum implementation which supports external calls.

\subsubsection{The Standard Oracle Way}
\label{sec:stdoracle}
A standard way to encode this use-case in Solidity is provided in Listing~\ref{lst:traditional}. Code relating to tracking of Ether sent in with a bet and the ability to withdraw winnings is not provided for the sake of simplicity. Given that a punter must first place a bet without the  random number being revealed to the punter, the process must be split up into two transactions. 

The first transaction involves a call by the punter to \texttt{beginBetOracle} which, beyond requiring payment which is left out here, needs to keep track of the bet placed and that a response is expected from the oracle to close the bet. The random number generator oracle needs to be notified to provide input which is communicated through an event (\texttt{BetPlaced}) emitted and logged to the blockchain on line 16.

\begin{lstlisting}[caption={Implementation of a betting smart contract that makes use of an external RNG oracle using a standard Ethereum network.},
                    label={lst:traditional},
                    language=solidity ]
contract BettingUsingStandardOracle {
    mapping(address => mapping(uint => bool)) public pending;
    mapping(address => int) public winnings;

    address public oracleAddress;    
    uint public pendingOracleNr;
    
    event BetPlaced(address indexed _from, uint oracleRefNr);
    
    constructor(address oracle) {
        oracleAddress = oracle;
    }
    
    function beginBetOracle() public {
        pending[msg.sender][pendingOracleNr] = true;
        emit BetPlaced(msg.sender, pendingOracleNr);
        pendingOracleNr++;
    }
    
    function continueBetOracle(address forPunter, uint oracleRef, bool punterWon) public returns(bool) {
        require(msg.sender == oracleAddress, "You are not the oracle!");
        bool isPending = pending[forPunter][oracleRef];
        if (isPending) {
            if (punterWon) {
                winnings[forPunter] += 1;
            } 
            pending[forPunter][oracleRef] = false;
            return true;
        }
        return false;
    }
}
\end{lstlisting}

Once the oracle comes across the logged event written to the blockchain, it will then send in its outcome as to whether the user has won (\texttt{punterWon}) for the specific bet by making a call to \texttt{continueBetOracle}. Based on whether or not the oracle reported that that user has won, the count of how many times the user won (\texttt{winnings}) is updated.

% \gp{Winnings not decremented if the player loses, since it is the number of times the player won.}

\subsubsection{The Active External Call Way}

Listing~\ref{lst:excall} provides the use-case smart contract code for the prototype external call implementation. A punter would initiate a transaction to call \texttt{betEXCALL} (along with their bet). The actual external call is invoked in line 7. In the prototype implemented for the scope of this paper, the Ethereum Virtual Machine (EVM) was altered to invoke an external call when a Solidity string assignment (or more specifically a \texttt{PUSH32} EVM instruction) is executed where the string is exactly 32 bytes and starts with ``http''. We leave language design issues pertaining to how this should be exposed to smart contracts to future work.

\begin{lstlisting}[caption={Implementation of a betting smart contract that makes use of an external RNG oracle using a standard Ethereum network.},
                    label={lst:excall},
                    language=solidity ]
contract BettingUsingEXCALL {
    mapping(address => int) public winnings;
    
    bytes public excall;
    
    function betEXCALL() public {
        excall = "http://localhost:8080/excallrand";
        if (excall[0] == '1') {
            winnings[msg.sender] += 1;
        } 
    }
}
\end{lstlisting}

The response from the external call is stored in \texttt{excall} and then depending upon the response (from the external service provider which responds with a Boolean value) \texttt{winnings} updated appropriately.

\subsection{Code complexity}

\subsubsection{Lines of Code}
Just by glancing at the two different versions of code it is immediately obvious that a platform that supports active external calls will result in simpler smart contract code. The standard approach implementation results in 34 lines of code; whilst the implemented active external call approach requires 14 lines of code --- around 40\% of that of the standard implementation. Indeed, however this is just an initial indication since further work on programming language design and programming language evaluation would need to be undertaken. However, it is clear that such an approach would result in less code. This is advantageous since less code likely means less potential for bugs~\cite{loc} --- and given that a single bug in a smart contract could result in extremely large financial losses~\cite{losses}, then an approach that could reduce such risks is desirable.

\subsubsection{Asynchronous and Synchronous Programming Models}
The main contributing factor for such an approach to require less lines of code is that it allows for, what would traditionally result in an asynchronous long-lived transaction across multiple invocations to a smart contract, to be programmed within a single synchronous code block. As opposed to an asynchronous model, a synchronous code block does not require for state to be stored across different transactions and therefore associated code can be removed.

\subsection{Throughput}
Transaction throughput and the ability for a blockchain to keep up with incoming transactions is vital to ensure that: (i) gas cost prices are kept low; (ii) user experience is not degraded by inducing long delays; and (iii) dApps can meet any hard deadlines or timing requirements. Therefore, it is crucial to evaluate the approach to see how it would affect transaction throughput. First, details pertaining to the experimental setup will be discussed followed by a presentation of results heeded.

\subsubsection{Experimental Setup}
To evaluate the approach, a private Ethereum `clique' proof-of-authority network was set up using the suggested block period of 15 seconds\footnote{\url{https://eips.ethereum.org/EIPS/eip-225}}. The same altered golang geth version (discussed in Section~\ref{sec:impl}) was used to run both implementations --- the standard oracle approach which makes use of standard EVM instructions, and the active external call approach which uses the \texttt{EXCALL} virtual machine instruction that was added in this prototype to demonstrate external calls.

\begin{figure}
\centering
\includegraphics[width=0.7\textwidth]{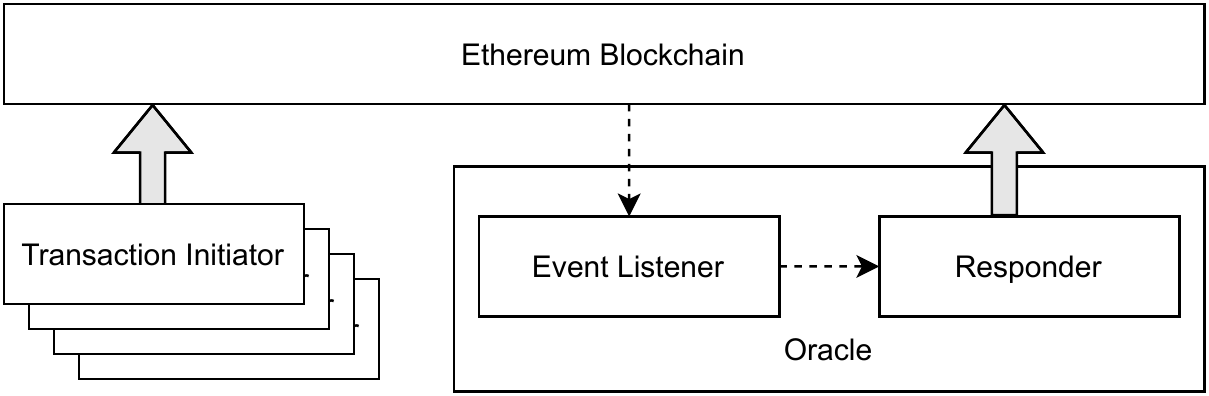}
\caption{The experimental setup for a standard oracle implementation for the given use-case.}
\label{fig:oraclesetup}
\end{figure}

\begin{figure}
\centering
\includegraphics[width=0.7\textwidth]{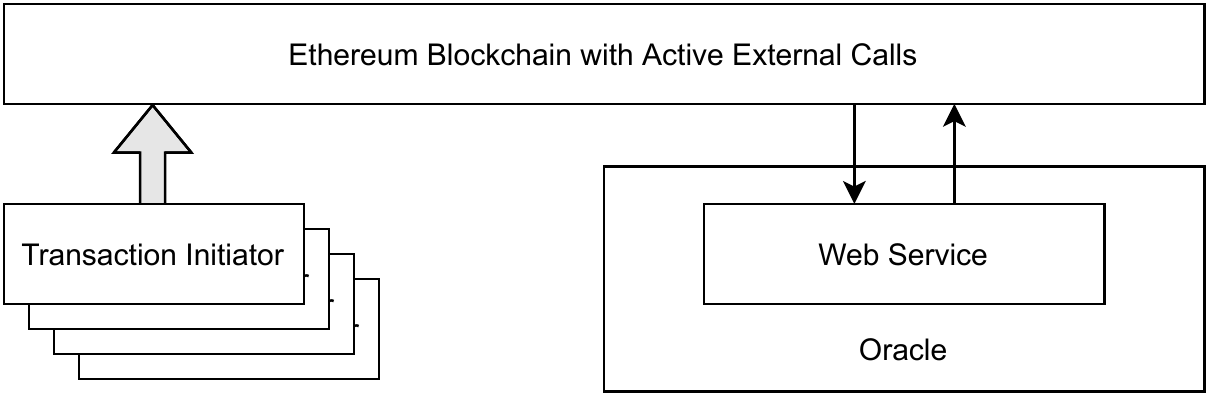}
\caption{The experimental setup for an active external call implementation for the given use-case.}
\label{fig:activesetup}
\end{figure}

The two implementations interact with the external world differently with: (i) the standard oracle approach, depicted in Figure~\ref{fig:oraclesetup}, using a passive method of logging data to the blockchain which requires an external oracle to listen for events written to the ledger, and thereafter can respond by invoking a smart contract function (in a subsequent block); and (ii) the active external call approach, depicted in Figure~\ref{fig:activesetup} which can make a direct external call to the oracle web service and immediately receive a response back.

Transaction initiators, representing punters in the particular use-case processes, invoke the respective smart contract to start the betting process (i.e. either the smart contract from Listing~\ref{lst:traditional} or Listing~\ref{lst:excall}). Each transaction initiator executes as a separate thread of execution (implemented as go routines) --- simulating simultaneous users. The grey arrows represent smart contract invocation transactions. The dashed lines represent passive internal process communication, whilst the two sold lines represent an HTTP Request/Response pair.

Experiments were conducted with varying: (i) concurrent transaction initiators (go routines) --- from 1 to 4; and (ii) the  number of times that each transaction initiator places a bet (10, 100 and 1000). An experiment run was deemed to be finished once all smart contract transactions were accepted and processed --- for the standard oracle implementation, this includes both transaction initiator smart contract function calls as well as the oracle's response smart contract call, whilst the active external call implementation only requires for the transaction initiators smart contract calls to be processed (since the external call takes place as part of the smart contract call invocation). Each experiment configuration was executed 4 times and the lower bound and upper bound timings recorded.

\begin{figure}
\centering
\includegraphics[width=\textwidth]{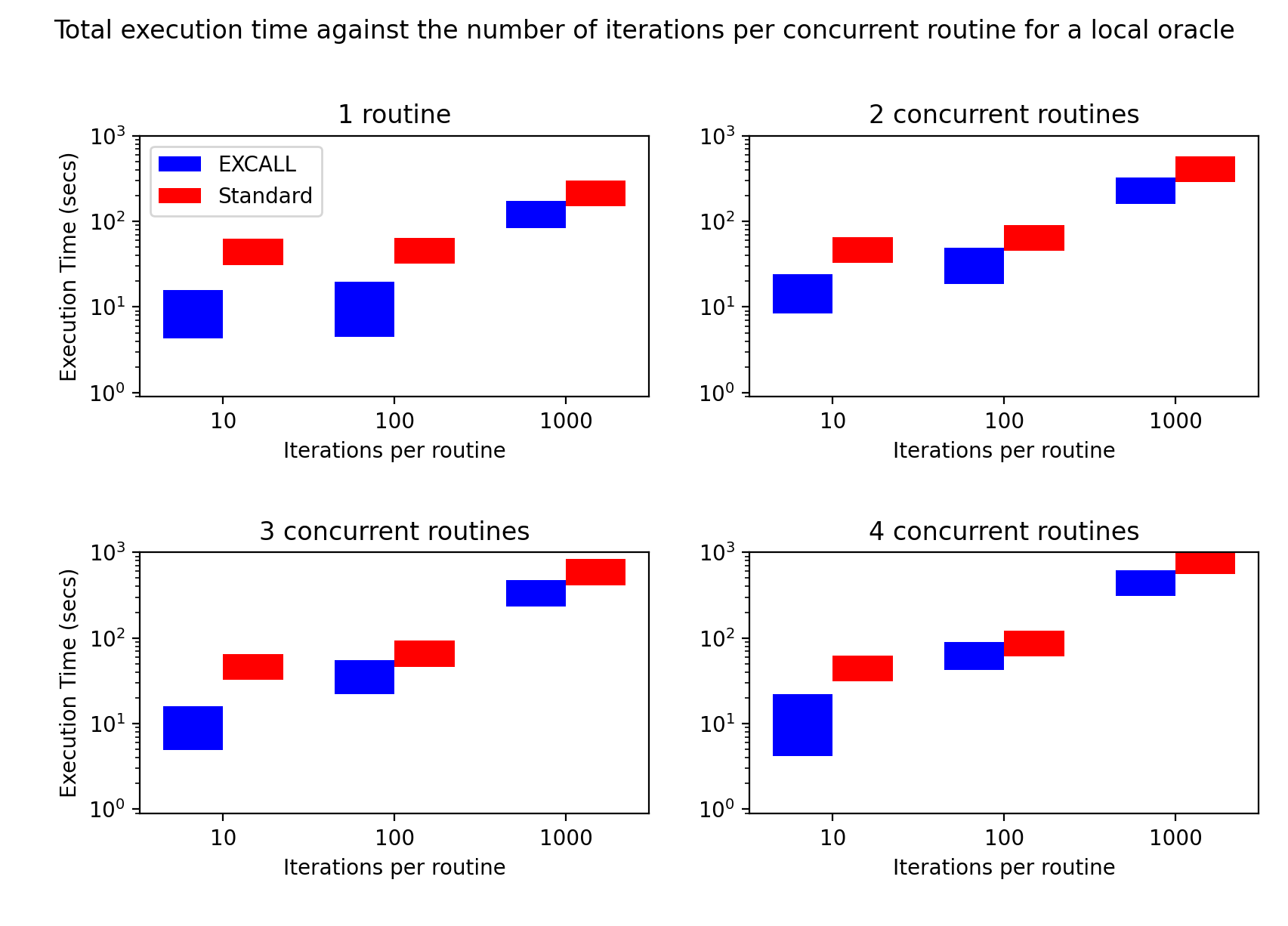}
\caption{Comparing execution time for processing use-cases which following user initiation requires oracle input before finalising a respective process. A local server was set up to act as the oracle data provider.}
\label{fig:resloc}
\end{figure}

\subsubsection{Results for local oracle service}
Experimentation was first conducted with an oracle service that is capable of both listening for events (in the case of the standard oracle approach) as well generating the response (i.e. the random number) itself. This oracle service was running on the same machine as the validator which also serves as a means of analysing performance overheads without inducing additional affects due to external network issues. The results are presented in Figure~\ref{fig:resloc}. As can be seen from the results, the active external call approach (EXCALL) consistently outperforms the standard oracle approach --- ranging from taking between: (i) 17\% to 60\% of the time when a single transaction initiator go routine was run; (ii) 41\% to 56\% of the time for 2 concurrent transaction initiators; (iii) 20\% to 72\% of the time to 3 concurrent transaction initiators; and (iv) 25\% to 85\% of the time for 4 concurrent initiators. Since the lower 10 and 100 iteration experiments finish between 4 seconds and 61 seconds for which only around 4 blocks would have been finalised, depending upon when the experiment started with relation to the last block the results could vary substantially given the 15 second block time.  Looking specifically at the 1,000 iteration experiments the EXCALL implementation results in taking between 56\% to 61\% of the time of the standard oracle approach.

\begin{figure}
\centering
\includegraphics[width=\textwidth]{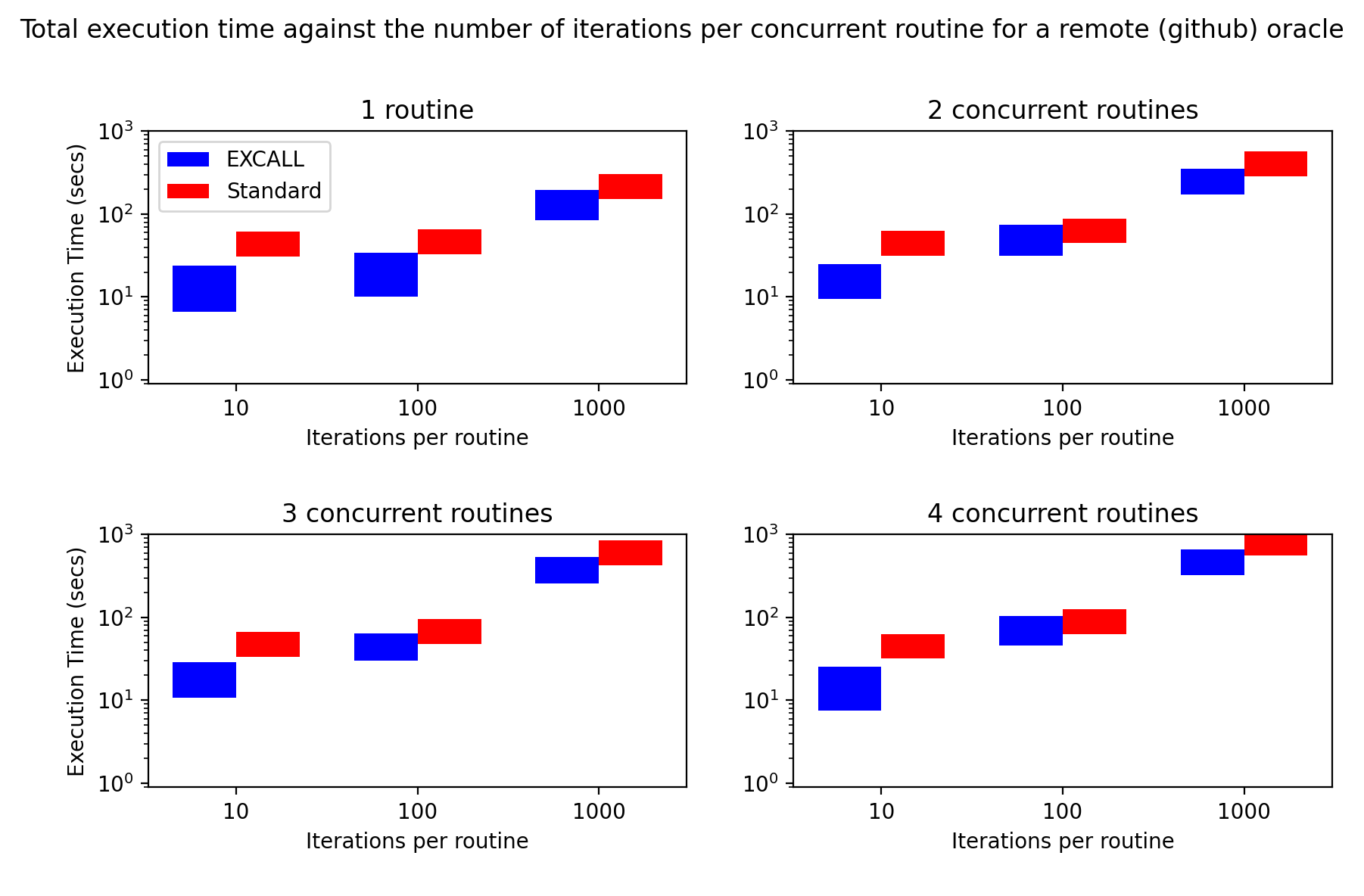}
\caption{Comparing execution time for processing a use-cases which following user initiation requires oracle input before finalising a respective process. A github resource was set up to act as the oracle data provider.}
\label{fig:resgit}
\end{figure}

\subsubsection{Results for retrieving web service response}

An external resource on github was used to determine a random number to decide whether the punter won. In the case of traditional oracles, since the service does not directly listen to events on a blockchain to act as an oracle and provide its data when required, an oracle service was also set up to act as an in-between: retrieving request events from the blockchain, fetching data from the service, and submitting it as an oracle.  In the case of active blockchain external calls, they could directly make the request to the github resource directly. 

% web services which reply with signed responses, 

% The difference from above is that the standard oracle local server will listen for events and then make a request to the external resource on github, whilst the active external call approach can make a request to the github resource directly. 

% in future (as may be a possibility based on works in progress as mentioned above\footnote{\url{https://wicg.github.io/webpackage/draft-yasskin-http-origin-signed-responses.html} \\ \url{https://developers.google.com/web/updates/2018/11/signed-exchanges}}) 
% then an active external call could be made to the external resource directly from the blockchain.

Results are depicted in Figure~\ref{fig:resgit}. Looking specifically at the results for experiments making 1,000 iterations, the active external call approach takes between 59\% to 74\% of the time it would take for the traditional oracle approach.

\subsection{Evaluation Recap}
Whilst this paper's contribution was to primarily show that active external calls from blockchain systems are in fact possible, contrary to the general consensus in the field, these results also demonstrate some potential advantages including: (i) allowing for less complexity when it comes to smart contracts that require external oracle input (which may lead to less bugs); and (ii) for the type of use-case presented herein higher smart contract transaction throughout would be able to be supported.

\section{Conclusions}
\label{sec:conc}
% \todo{
Despite that it is widely accepted that blockchain systems (particularly public unpermissioned ones) cannot execute calls to external systems due to the requirement for computation to reach a deterministic state and to avoid denial-of-service attacks on the service provider, we have shown a way of enabling such external calls initiated from the blockchain/DLT, without introducing new points-of-trust, and reducing overheads. 

This work opens various new avenues for research. We are currently looking into a deeper analysis of comparative gas costs between the two approaches, and also at language-design issues to cater for external calls in smart contract programming. Another challenge is having trust annotations and analysis at the programming language level, which becomes much more straightforward given that external calls are explicitly annotated. From an application perspective, the scope of possibilities is widened and we are also looking at use cases whose description becomes substantially easier with the possibility of external calls.

% Contrary to the general consensus, in this paper, we have demonstrated a method for Blockchain and DLT systems that allows for direct external calls to be initiated from the Blockchain/DLT itself (or even from a transaction initiator's software if suitable). 

% Although it is often said that blockchain-based computation needs to be deterministic~\cite{sankar2017survey}, it is important to highlight that this requirement does not mean that different computation cannot take place to reach that state. 
% }

% \todo{
A prototype demonstrating verifiable external calls has been implemented and available from \url{https://github.com/joshuaellul/excalls}, and further information on the project is available from: \url{https://joshuaellul.github.io/excalls/}.
% }

\bibliographystyle{ACM-Reference-Format}
\bibliography{sample-base}

\end{document}

%% file: solidity.tex
\usepackage{listings, xcolor}

\definecolor{verylightgray}{rgb}{.97,.97,.97}

\lstdefinelanguage{solidity}{
	keywords=[1]{static, object, byte, bool, string, anonymous, assembly, assert, balance, break, call, callcode, case, catch, class, constant, continue, contract, debugger, default, delegatecall, delete, do, else, emit, event, export, external, false, finally, for, function, gas, if, implements, import, in, indexed, instanceof, interface, internal, is, length, library, log0, log1, log2, log3, log4, memory, modifier, new, payable, pragma, private, protected, public, pure, push, require, return, returns, revert, selfdestruct, send, struct, suicide, super, switch, then, this, throw, transfer, true, try, typeof, using, value, view, while, with, addmod, ecrecover, keccak256, mulmod, ripemd160, sha256, sha3}, % generic keywords including crypto operations
	keywordstyle=[1]\color{blue}\bfseries,
	keywords=[2]{Runtime, Storage, address, Features, MonitorInvariant, SmartContract, InvalidOperationException, bytes, bytes1, bytes2, bytes3, bytes4, bytes5, bytes6, bytes7, bytes8, bytes9, bytes10, bytes11, bytes12, bytes13, bytes14, bytes15, bytes16, bytes17, bytes18, bytes19, bytes20, bytes21, bytes22, bytes23, bytes24, bytes25, bytes26, bytes27, bytes28, bytes29, bytes30, bytes31, bytes32, enum, int, int8, int16, int24, int32, int40, int48, int56, int64, int72, int80, int88, int96, int104, int112, int120, int128, int136, int144, int152, int160, int168, int176, int184, int192, int200, int208, int216, int224, int232, int240, int248, int256, mapping, uint, uint8, uint16, uint24, uint32, uint40, uint48, uint56, uint64, uint72, uint80, uint88, uint96, uint104, uint112, uint120, uint128, uint136, uint144, uint152, uint160, uint168, uint176, uint184, uint192, uint200, uint208, uint216, uint224, uint232, uint240, uint248, uint256, var, void, ether, finney, szabo, wei, days, hours, minutes, seconds, weeks, years},	% types; money and time units
	keywordstyle=[2]\color{teal}\bfseries,
	keywords=[3]{ContractFeatures, TriggerType, Main, block, blockhash, coinbase, difficulty, gaslimit, number, timestamp, msg, data, gas, sender, sig, value, now, tx, gasprice, origin},	% environment variables
	keywordstyle=[3]\color{violet}\bfseries,
	identifierstyle=\color{black},
	sensitive=false,
	comment=[l]{//},
	morecomment=[s]{/*}{*/},
	commentstyle=\color{gray}\ttfamily,
	stringstyle=\color{red}\ttfamily,
	morestring=[b]',
	morestring=[b]"
}